\documentclass[12pt]{article}
\usepackage{epsfig,amsfonts,amssymb}
\usepackage{hyperref}

\usepackage{cite}
\usepackage{cite}

\def\ZZZ{{\hbox{ Z\kern-1.6mm Z}}}
\def\RRR{{\hbox{ R\kern-2.4mm R}}}
\def\CCC{{\hbox{ C\kern-2.0mm C}}}
\def\zzz{{\hbox{z\kern-1mm z}}}

\newcommand{\qeq}{{\hbox{=\kern-2.3mm ? \kern.5mm }}}
\renewcommand{\qeq}{=}

\newcommand{\eps}{\epsilon}

\newcommand{\vp}{\varphi}

\newcommand{\II}{{\cal I}}

\newcommand{\JJ}{{\cal J}}

\newcommand{\OO}{{\cal O}}

\newcommand{\wt}{\widetilde}

\newcommand{\NN}{{\cal N}}

\newcommand{\be}{\begin{equation}}
\newcommand{\ee}{\end{equation}}
\newcommand{\ben}{\begin{eqnarray}\displaystyle}
\newcommand{\een}{\end{eqnarray}}

\newcommand{\refb}[1]{(\ref{#1})}
\newcommand{\p}{\partial}
\newcommand{\sectiono}[1]{\section{#1}\setcounter{equation}{0}}

\def\one{{\hbox{ 1\kern-.8mm l}}}
\def\zero{{\hbox{ 0\kern-1.5mm 0}}}

\newcommand{\bea}[1]{\begin{eqnarray}\label{#1} }
\newcommand{\eea}{\end{eqnarray}}

\newcommand{\eqref}{\refb}

\usepackage{bm}
\usepackage[table]{xcolor}

\begin{document}

\baselineskip 24pt

\begin{center}

{\Large \bf D-instanton Perturbation Theory}


\end{center}

\vskip .6cm
\medskip

\vspace*{4.0ex}

\baselineskip=18pt

\centerline{\large \rm Ashoke Sen}

\vspace*{4.0ex}

\centerline{\large \it Harish-Chandra Research Institute, HBNI}
\centerline{\large \it  Chhatnag Road, Jhusi,
Allahabad 211019, India}


\vspace*{1.0ex}
\centerline{\small E-mail:  sen@hri.res.in}

\vspace*{5.0ex}

\centerline{\bf Abstract} \bigskip

D-instanton world-volume theory has open string zero modes describing collective coordinates
of the instanton. The usual perturbative amplitudes in the D-instanton background
suffer from infra-red divergences due to the
presence of these  zero modes, and  the usual approach of analytic continuation in
momenta does not work since all open string states on a D-instanton carry strictly zero
momentum.  String field theory is well-suited for tackling 
these issues. However
we find a new subtlety due to the existence of additional zero modes in the ghost
sector. This causes a breakdown of the Siegel gauge, but a different gauge fixing 
consistent with the BV formalism
renders the perturbation theory finite and unambiguous. At each order, this produces 
extra contribution to the amplitude besides what is obtained from integration over the moduli
space of Riemann surfaces.

\vfill \eject

\baselineskip 18pt

\tableofcontents

\sectiono{Introduction} \label{sintro}

String theory is usually formulated using the world-sheet approach. This expresses 
all perturbative
amplitudes in string theory as integrals over the moduli spaces of Riemann surfaces 
with punctures, with the integrands computed in terms of appropriate correlation functions
in the world-sheet 
conformal field theory of matter and ghost fields. However
the integrands are often singular at the boundaries of the moduli spaces, leading to singular
integrals. In many cases one can nevertheless define the integral by analytic continuation in
the external momenta. However in some cases, involving mass renormalization and vacuum
shift, analytic continuation in external momenta is not enough to remove the divergences. 
In these cases we need to 
use string field theory to get well defined finite answers for all physical
quantities\cite{1703.06410}.

The problem becomes particularly acute in the presence of D-instantons\footnote{D-instantons
have been recently explored in string field theory in a different context -- as classical solution
on multiple D3-branes\cite{1902.10955,1910.00538}. 
Our goal here is to study perturbation theory in the presence of
D-instantons.}
-- D-branes with
Dirichlet boundary condition along all non-compact
directions including (euclidean) time, since open strings
living on such D-branes do not carry any momenta and therefore the
divergences cannot be
removed by analytic continuation in external momenta. Often one can give physical 
arguments as to why the divergences cancel\cite{9407031,9701093,0211250}; 
however since this requires combining 
different amplitudes, after cancelling divergences we are left with a finite ambiguity that
cannot be fixed. A particular example of this arose in a recent analysis of D-instanton 
contribution to two dimensional string theory\cite{1907.07688}. However, 
since string field theory is a regular ultra-violet finite quantum field theory
with well defined action (up to field redefinition) we do not expect any ambiguity to arise in
computation of amplitudes in string field theory.  
Indeed, one such ambiguity in the two dimensional string theory was 
eventually resolved
using string field theory, leading to results in agreement with those in the dual matrix 
model\cite{1908.02782}.

The divergences in the world-sheet theory in the presence of D-instantons arise from
various sources. The first source comprises 
the collective coordinates of the D-instanton associated with
the freedom of translating the D-instanton along the space-time directions transverse
to the brane. These collective coordinates correspond to zero modes in string field 
theory\footnote{Note that the relevant part of string field theory, describing open strings living
on the D-instanton, is a zero dimensional field theory. Therefore we shall use the
words mode and field interchangeably.} --
modes with vanishing quadratic term in the action. Therefore the
propagator diverges, leading to divergences in the perturbative amplitude.
In the world-sheet description, these show up as logarithmic divergences in the
integral over the moduli spaces of Riemann surfaces with punctures.
While the conventional world-sheet approach does not give us a systematic procedure for
dealing with these divergences, 
the treatment of these collective modes in string field theory is the same
as in ordinary quantum field theory. 
Instead of treating these modes perturbatively, we leave them
unintegrated at the beginning, evaluate the Feynman diagrams using the propagators of the
other modes, and after summing over all the Feynman diagrams we integrate over the
collective modes. This is expected to recover the energy-momentum conserving delta
function which is initially absent in the presence of D-instantons, since space-time translation
invariance is broken. This gives an unambiguous procedure for treating the divergence 
associated with the collective modes. Indeed this treatment of the collective modes was
used in \cite{1908.02782} 
to fix a constant in two dimensional string theory that remains ambiguous in
the usual word-sheet approach.

The second source of divergence in the perturbative amplitudes, expressed as
integrals over moduli spaces of punctured Riemann surfaces, can be traced to
open string tachyons if they are present.
Normally theories with tachyons are not sensible, unless we can find a new
field configuration where tachyons are absent, but D-instantons are different in this
respect. The presence of tachyonic open string state on a D-instanton implies that the
D-instanton represents a saddle point of the action and therefore the weight factor
$e^S$ in the path integral has a local extremum instead of a local maximum at the solution. 
Nevertheless
such instantons may give sensible contribution to the path integral, as was convincingly
demonstrated in the recent analysis in two dimensional string 
theory\cite{1907.07688,1912.07170}. In fact,
D-instantons with tachyonic mode may be present even in supersymmetric
string theories,  {\it e.g.} the non-BPS
D-instanton in the type IIA string theory. From the
point of view of string field theory, since the open string modes do not carry momentum,
there is no difficulty in carrying out perturbation theory with tachyons -- the propagator of
a mode of mass $m$ is given by $1/m^2$ irrespective of whether $m^2$ is positive
or negative. However its world-sheet representation, where we represent $1/m^2$ as
$\int_0^\infty ds \, e^{-m^2\, s}$, diverges for $m^2<0$.  
Therefore, if instead of using the world-sheet representation of the amplitude
we use the string field theory representation, there is no divergence in perturbation 
theory. This has been 
discussed extensively in \cite{1902.00263,1702.06489}. 

There is a third source of divergence that will be the main focus of this paper. This is due to
the presence of additional open string zero modes on the D-instanton that are not associated
with the collective coordinates of the D-instanton. For D-instantons in bosonic string theory,
these are associated with the pair of states $|0\rangle$ and $c_1c_{-1}|0\rangle$. These
states satisfy the Siegel
gauge condition $b_0|\Psi\rangle=0$ that is normally used in string field theory, but the
associated fields have
vanishing kinetic term. Therefore the propagators associated with these modes are
infinite and perturbative amplitudes diverge. Furthermore, in this case we cannot
remove these divergences by treating them as collective modes. The remedy turns out to
be to alter the gauge fixing procedure in the zero mode sector -- instead of using the
gauge fixed action we use the original gauge invariant action in this sector. 
Of course this can
not be done in an ad hoc fashion, but we show that the Batalin-Vilkovisky
(BV) 
formalism\cite{bv,bv1,henn}, that underlies the formulation of 
open-closed string field theory\cite{zwiebach_open,9705241,1907.10632}, 
allows us to do this. The net
effect of this is that instead of using the states $|0\rangle$ and $c_1c_{-1}|0\rangle$ in the
expansion of the string field, we need to use the states $|0\rangle$ and $c_0|0\rangle$ in the
expansion.
This leads to well defined perturbation
expansion without any divergent propagator.

The rest of the paper is organized as follows. In \S\ref{scomb} we review the organization
of the terms in D-instanton perturbation theory. In particular we discuss why we must 
include in our analysis certain class of disconnected diagrams but exclude other classes
of disconnected diagrams. In \S\ref{sdiv} we discuss various types of divergences that
arise in perturbation theory in the presence of D-instantons and their remedy.
In particular, \S\ref{stach} discusses the divergences due to the collective coordinates
and open string tachyons, 
\S\ref{sghost} discusses the divergences due to the zero modes from the
ghost sector, and \S\ref{ssum} contains a summary of the algorithm needed to tackle all the divergences
systematically.
In \S\ref{swarm} we demonstrate the need for this new treatment of the ghost
zero modes by analyzing a specific amplitude -- a disk amplitude with four external collective
modes of the open string. We show that in order to get the correct result, we must include
the contribution of the out of Siegel gauge mode, associated with the state $c_0|0\rangle$,
in the computation. In \S\ref{ssuper} we discuss similar issues for the ghost sector 
zero modes for D-instantons in superstring theory.

\sectiono{Diagrammatics of D-instanton contribution} \label{scomb}

Let us consider a quantum field theory with instanton solutions. 
In order to identify the instanton contribution to the
Green's function of a collection of operators, which we shall denote by $\OO$, we shall divide the path integral over
the fields $\Phi$ into different sectors labelled by their instanton number. 
For simplicity we shall analyze the contribution up to one instanton sector, but the analysis
can be easily generalized to the multi-instanton sector. We
denote by $\Phi_p$ the fluctuations around the vacuum solution and by $\Phi_I$ the fluctuations around the single
instanton solution, and express the correlation function of $\OO$ as\footnote{Throughout the paper we shall use the 
convention that 
the action $S$ appears in
the integrand of Euclidean path integral as $e^S$.}
\be\label{er1}
\langle \OO\rangle = {\int D\Phi_p \exp[S_p] \, \OO + \NN\, e^{-C/g_s} \int D\Phi_I \exp[S_I] \, \OO\over
\int D\Phi_p \exp[S_p] + \NN\, e^{-C/g_s} \int D\Phi_I \exp[S_I]}\, ,
\ee
where $-C/g_s$ is the instanton action and 
$\NN$ is a normalization constant that gives the ratio of the integration measure in the instanton sector
and in the perturbative sector. $S_p$ denotes the action of the fluctuating fields $\Phi_p$ 
around the vacuum solution and $S_I$ denotes the action of the fluctuating fields 
$\Phi_I$ 
around the one instanton solution.

If the instanton under consideration represents a D-instanton in string theory,
then the various terms in this expansion have 
clear interpretation. $\int D\Phi_p \exp[S_p] \, \OO$ gives the
amplitudes containing world-sheets that do not have any boundary ending on the D-instanton,
but we must allow world-sheets with multiple disconnected components, 
including vacuum bubbles
which do not have any external vertex operator insertion. $\int D\Phi_I \exp[S_I] \, \OO$ 
gives the
perturbative amplitudes containing world-sheets that may have
multiple disconnected components, possibly including vacuum bubbles, but at least one of the world-sheets
must have at least one boundary ending on the D-instanton. The factors in the denominator have similar 
interpretation, except that there is no external vertex operator insertion.

Keeping terms containing at most one power of $e^{-C/g_s}$, we can expand \refb{er1} as
\be\label{er2}
\langle \OO\rangle = {\int D\Phi_p \exp[S_p] \OO\over
\int D\Phi_p \exp[S_p]} + \NN\, e^{-C/g_s}  { \int D\Phi_I \exp[S_I] \OO\over
\int D\Phi_p \exp[S_p]} - \NN\, e^{-C/g_s}  {\int D\Phi_p \exp[S_p] \OO\over
\int D\Phi_p \exp[S_p]} { \int D\Phi_I \exp[S_I] \over
\int D\Phi_p \exp[S_p]}\, .
\ee
We can now interpret the various terms in string theory as follows.
\begin{enumerate}
\item
The first term is the perturbative amplitude, possibly containing disconnected world-sheets but there should
be no boundary ending on D-instanton. 
The division by the denominator removes from this all factors
containing disconnected bubbles. However, disconnected world-sheets are still allowed as long as each
component has at least one vertex operator insertion.
\item
The second term represents amplitudes in the instanton background, but 
the division by the denominator removes
all factors containing disconnected bubbles {\em in the perturbative amplitude}. Note that we do not
remove bubbles in the instanton background.  For D-instantons this means that we 
sum over world-sheets for which each connected component has either
insertion of an external
vertex operator, or a boundary ending on the D-instanton, or both.
\item The third term is a subtraction term containing product of two factors. The first one represents
the perturbative amplitude with the bubble diagrams removed. The second term represents vacuum
bubble diagrams in the presence of the instanton, but containing no factors with perturbative vacuum
bubble. For D-instantons this means that we must remove all diagrams in which all the external state
vertex operators end on world-sheets without any boundary ending on the D-instanton, even if they are
multiplied by vacuum bubbles containing boundaries that do end on the D-instanton. 
\end{enumerate}
Therefore the rules for computing a single 
D-instanton contribution to a given amplitude is to sum over all
world-sheet diagrams, possibly containing disconnected components, but 
subject to the following conditions:
\begin{enumerate}
\item Each of these disconnected components must have either at least one 
boundary ending on the
D-instanton or at least one closed string vertex operator. 
\item At least one of the disconnected 
components must have both, a boundary ending on the D-instanton and a closed string vertex
operator insertion. 
\end{enumerate}
Each such contribution will be multiplied by a single factor of $\NN\, e^{-1/g_s}$,
irrespective of the number of disconnected components it has.

\sectiono{Dealing with divergences} \label{sdiv}

We shall use string field theory to evaluate the D-instanton contribution to the
physical amplitudes. 
As will be explained shortly, this is needed to
deal with infrared divergences. 
The string field theory that is relevant for this problem is the interacting 
field theory of open and closed strings,
with open strings satisfying boundary conditions associated with the D-instanton. The collection of open
and closed string fields together correspond to the set of fields $\Phi_I$ in \refb{er1},\refb{er2}, with the open
strings describing modes that are localized on the D-instanton and closed strings describing modes that are
not localized on the instanton. In contrast, the modes $\Phi_p$ with action $S_p$
in  \refb{er1},\refb{er2} are 
described by closed string field theory without any D-instanton background.

In any amplitude, 
the external states of interest will be closed strings (or in general situation
open strings living on D-branes other than D-instantons) -- the open strings living
on the transient D-instantons do not correspond to asymptotic states. 
However, a subset of the open string fields represent 
the collective coordinates of the D-instanton,
associated with translation along
space-time directions, and we cannot carry out the usual perturbation theory in which these zero
modes propagate in the internal state, -- they have divergent propagator. 
Therefore in the path integral over the string fields $\Phi_I$ in \refb{er2}, 
we must leave these zero modes unintegrated while integrating over
all other open string fields, and carry out 
integration over these zero modes at the very end. 
In perturbation theory, this means that
we must subtract these zero mode contributions from
the internal open string propagators, allow arbitrary number of these zero modes to appear as
external states together with the closed string states, sum over all Feynman diagrams and all
possible number of insertions of the zero mode fields $\phi$ in amplitudes with a given set of external
closed string states, 
and at the end integrate over these zero modes $\phi$
explicitly. On physical grounds, these zero mode integrals are expected to restore the
space-time momentum conserving delta functions that are otherwise missing in the amplitudes
in the presence of D-instantons. In the following we shall discuss the systematic procedure for
doing this analysis in string field theory.\footnote{The only ambiguity that does not seem to
be resolved in the current formulation of 
string field theory is the overall normalization constant $\NN$ in \refb{er1}, \refb{er2}. 
This is related to
the freedom of adding a constant to the string field theory action around the D-instanton.}

\subsection{Tachyons and collective modes} \label{stach}

The world-sheet expressions for the amplitudes in string theory 
often diverge from the region where certain
number of vertex operators come together, or, more generally, when a Riemann
surface with punctures
degenerates. Since the divergences of interest to us will arise from integration over the open string
fields, we shall focus exclusively on these -- divergences associated with closed strings, if present,
can be dealt with by following the procedure described in \cite{1902.00263}.
The origin of these divergences in string field theory 
can be understood by
noting that the world-sheet approach replaces the $1/L_0$ factor in the Siegel
gauge open string propagator by:
\be\label{eprop}
1/L_0 \to \int_0^1 dq \, q^{L_0-1}\, .
\ee
This is an identity for $L_0>0$ but fails for $L_0\le 0$. For $L_0<0$ the left hand side is well defined but the right
hand side is divergent. The world-sheet description of the amplitude uses the right hand side and is therefore
divergent, while string field theory uses the left hand side and gives 
a finite result. Therefore such divergences in the world-sheet amplitude may be dealt with
simply by suitably parametrizing the moduli space of Riemann surfaces near degeneration
points by variables induced from string field theory, including $q$, 
and then replacing integrals of the form $\int_0^1 dq\, q^{\beta-1}$
by $1/\beta$ for $\beta<0$.

For $L_0=0$ both sides diverge. This is a reflection of the
presence of zero mode(s) in the open string sector.
While the world-sheet approach does not provide us with a systematic way of dealing with
these  divergences, in string field theory typically the zero modes would have definite
interpretation and therefore there is an unambiguous procedure for dealing with them.
In this subsection we shall describe the procedure for dealing with one set of these zero modes -- those
associated with the collective coordinates of the D-instanton. We shall denote these zero
modes collectively by $\phi$. As already mentioned, 
the solution string field theory offers for dealing with such zero modes is to first 
carry out the path integral over all string fields other than $\phi$,  
for fixed background $\phi$, and then carry out the integration
over $\phi$ explicitly. In the world-sheet computation, this translates to the following
algorithm\cite{1908.02782}:
\begin{enumerate}
\item Removing integration
over these zero modes in the path integral corresponds to removing the 
singular contributions due to these zero modes from the internal open string propagators of the
Feynman diagrams. 
In the world-sheet description, this requires parametrizing 
the moduli space of Riemann surfaces near degeneration
points by variables induced from string field theory, including $q$, and then
removing the singular contribution to the integral
proportional to $\int_0^1dq/q$ due to these zero modes. \label{pointone}
\item Since we are supposed to carry out the path integral with fixed background 
$\phi$, we have to
compute amplitudes with
external $\phi$ states (and closed string states) 
even though we are ultimately interested in amplitudes with external closed strings only.
Near each degeneration point we follow the subtraction scheme mentioned in 
point \ref{pointone}.
\item
After
computing the  relevant amplitude in background $\phi$, we carry out integration over 
$\phi$. This is expected
to restore momentum conservation that is broken in the presence of a single D-instanton.
For example if $\xi$ denotes the set of
collective coordinates associated with space-time 
translation and $p$ denotes the total momentum of external closed strings in an
amplitude, then the amplitude is expected to be proportional to
$e^{ip.\xi}$ so that integration over $\xi$ gives a factor of $\delta(p)$.
However, this may not be manifest, since the modes $\phi$ that 
arise from string field theory may
be related to the collective coordinates $\xi$
by a field redefinition. In that case, the easiest way to
see
the momentum conserving delta function arising out of the zero mode integration
will be to try to
use a specific version of string field theory in which the modes $\phi$ 
coming from string field theory
coincide with the collective coordinates $\xi$
without any field redefinition\cite{1908.02782}. 
In such cases
one recovers the momentum conserving delta function directly from the 
integration over the
zero modes $\phi$ arising in string field theory.
Alternatively, one could use a generic version of string field theory but find the
explicit field redefinition that relates the open string modes $\phi$ to the collective
coordinates $\xi$ that have the coupling proportional to
$e^{i p.\xi}$\cite{appear}. The Jacobian associated with this field redefinition will
have to be taken into account in the analyis.
After this one can 
carry out the integration over the $\xi$ 
modes and recover the momentum conserving delta
function. 
\end{enumerate}

Before concluding this subsection, we shall describe the vertex operator for the zero modes associated with
the collective coordinates.
Let us for definiteness, consider the zero mode associated with translation along the
(euclidean) time coordinate.
The unintegrated world-sheet vertex operator associated with the
corresponding open string state is given by 
$c\,\p X$ where $b,c$ denote the usual world-sheet ghost fields and 
$X$ is the world-sheet scalar labelling the time direction.  The zero modes associated with
translation along other directions can be described in a similar way.

\subsection{Ghost zero modes and the inadequacy of Siegel gauge} \label{sghost}

Collective coordinates are not the only open string zero modes in  
string field theory in the presence of a 
D-instanton -- there are other zero modes arising in the ghost sector that require different
treatment. In order to understand this we need to begin with a brief review of the
BV formalism\cite{bv,bv1,henn}.

In the BV formalism
for open-closed string field theory\cite{9705241,1907.10632},
we take a generic open string field $|\Psi_o\rangle$ or closed string field $|\Psi_c\rangle$
to be a state in the world-sheet CFT of arbitrary ghost number (subject to the condition
$b_0^-|\Psi_c\rangle=0=L_0^-|\Psi_c\rangle$ for closed string fields) and expand it as 
linear combination of a 
complete set of basis states.\footnote{We define $b_0^\pm=(b_0\pm \bar b_0)$, $L_0^\pm=(L_0\pm \bar L_0)$ and $c_0^\pm = (c_0\pm \bar c_0)/2$. Furthermore, we
assign ghost number 1 to $c$, $\bar c$ and ghost number $-1$ to $b$, $\bar b$.}
The coefficients of expansion are the dynamical variables
of the theory, with the coefficients of the open string states of ghost number
$\le 1$ and
closed string states of ghost number $\le 2$ considered as fields, and the coefficients of the
open string states of ghost number $\ge 2$ and closed string states of ghost number $\ge 3$
considered as anti-fields. Up to signs, the 
pairing between fields and anti-fields is done via BPZ inner
product, with an insertion of $c_0^-$ in the inner product of closed string states. For example,
if $|\vp_r\rangle$ denotes a basis of open string states of ghost number $\le 1$ and 
$|\vp^r\rangle$ is a
basis of open string states of ghost number $\ge 2$, satisfying the orthonormality condition
$\langle\vp^r|\vp_s\rangle=\delta^r_s$, and if we expand the open string field as
$\sum_r \left\{\psi_r |\vp^r\rangle + \psi^r |\vp_r\rangle\right\}$, then $\psi_r$ is the anti-field of 
$\psi^r$ up to a sign. Similarly if
$|\phi_r\rangle$ denotes a basis of closed string states of ghost number $\le 2$ and 
$|\phi^r\rangle$ is a
basis of closed string states of ghost number $\ge 3$, each annihilated by $b_0^-$ and
$L_0^-$, and satisfying the orthonormality condition
$\langle\phi^r|c_0^-|\phi_s\rangle=\delta^r_s$, and if we expand the closed string field as
$\sum_r \left\{\chi_r |\phi^r\rangle + \chi^r |\phi_r\rangle\right\}$, then $\chi_r$ is the anti-field of 
$\chi^r$ up to a sign. It is however possible to define new fields and anti-fields by making
a symplectic transformation that preserves the anti-bracket. Therefore if we introduce new
orthonormal basis $|\tilde\vp_r\rangle$ and $|\tilde\vp^r\rangle$ for open string states and
$|\tilde\phi_r\rangle$  and 
$|\tilde\phi^r\rangle$ for closed string states, with 
$\langle\tilde\vp^r|\tilde\vp_s\rangle=\delta^r_s$,
$\langle\tilde\vp^r|\tilde\vp^s\rangle=0$, $\langle\tilde\vp_r|\tilde\vp_s\rangle=0$ and 
$\langle\tilde\phi^r|c_0^-|\tilde\phi_s\rangle=\delta^r_s$,
$\langle\tilde\phi^r|c_0^-|\tilde\phi^s\rangle=0$, 
$\langle\tilde\phi_r|c_0^-|\tilde\phi_s\rangle=0$,
and expand the open string field as
$\sum_r \{\tilde\psi_r |\tilde\vp^r\rangle + \tilde\psi^r |\tilde\vp_r\rangle\}$ 
and the closed string field as
$\sum_r \{\tilde\chi_r |\tilde\phi^r\rangle + \tilde\chi^r |\tilde\phi_r\rangle\}$, then we can treat
$\tilde\psi^r$ and $\tilde\chi^r$ as fields and $\tilde\psi_r$ and $\tilde\chi_r$ as the
corresponding anti-fields up to sign.

In the BV formalism, the 
path integral of string field theory, weighted by the exponential of the action, 
is to be carried out over
a Lagrangian submanifold, which corresponds to setting the anti-fields to zero, possibly after
making a symplectic transformation.  The result of the path integral can be shown to be 
(formally) independent of the choice of the Lagrangian submanifold. If we use the original
definition of fields and anti-fields and define the Lagrangian submanifold to be the subspace
$\psi_r=0$, $\chi_r=0$, then the remaining open string fields have ghost number $\le 1$ and
the remaining closed string fields have ghost number $\le 2$. Ghost number conservation
then implies that the action depends only on the open string fields of ghost number 1 and
closed string fields of ghost number 2, i.e.\ the classical fields. The integration over the open
fields of ghost number $\le 0$ and closed string fields of ghost number $\le 1$ decouple for
physical amplitudes,
and effectively corresponds to division by the volume of the gauge group. The resulting
path integral can be identified as the conventional 
path integral over all the fields without any gauge fixing, since all the classical fields -- open
string fields of ghost number 1 and closed string fields of ghost number 2, are to be 
integrated
over. This is formally the correct path integral, but produces singular perturbation 
expansion,
since the gauge symmetry remains unfixed. In particular the kinetic operator will have zero
eigenvalues due to the presence of pure gauge states of the form 
$Q_B|s\rangle$.

On the other hand, if we choose to expand the string fields in the new basis 
$|\tilde\vp_r\rangle$, $|\tilde\vp^r\rangle$, 
$|\tilde\phi_r\rangle$  and 
$|\tilde\phi^r\rangle$, satisfying
\be
c_0|\tilde\vp^r\rangle=0, \quad b_0|\tilde\vp_r\rangle=0, \quad
c_0^+|\tilde\phi^r\rangle=0, \quad b_0^+|\tilde\phi_r\rangle=0,
\ee
and define the Lagrangian submanifold by setting $\tilde\psi_r$ and $\tilde\chi_r$ to zero,
then the remaining open string field $\sum_r \tilde\psi^r |\tilde\vp_r\rangle$ and
the closed string field $\sum_r \tilde\chi^r |\tilde\phi_r\rangle$ satisfy the Siegel gauge conditions
$b_0|\Psi_o\rangle=0$, $b_0^+|\Psi_c\rangle=0$. The
resulting path integral is now carried out over fields of all ghost numbers and corresponds
to the usual gauge fixed path integral, leading to well defined perturbation theory in a generic
open-closed string field theory. We shall see however that in the presence of D-instantons
this procedure leads to singular path integral.

For open string fields living on D-instantons, which do not carry
any momentum, special care is needed to deal with the
ghost excitations carrying $L_0=0$. For this let us consider the basis states 
$|0\rangle$, $c_0|0\rangle$, $c_1 c_{-1}|0\rangle$ and $c_1 c_0 c_{-1}|0\rangle$, and expand
the open string field in this sector as
\be\label{eexpand}
\psi^1 \, c_0|0\rangle + \psi^2|0\rangle+  \psi_1 \, c_1 c_{-1}|0\rangle 
+ \psi_2 \, c_1 c_0 c_{-1}|0\rangle\, .
\ee
In the original formulation, $\psi^1$ and $\psi^2$ are fields and $\psi_1$ and $\psi_2$ are
anti-fields. Therefore
the gauge invariant path integral will correspond to setting $\psi_1$ and $\psi_2$ to 0.
On the other hand
the Siegel gauge path integral will correspond to setting $\psi^1$ and $\psi_2$ to 0.
However in this case the quadratic term in the action, being proportional to $L_0$, 
does not depend on the remaining
fields $\psi^2$ and $\psi_1$ that multiply the $L_0=0$ states. This makes the path integral
over $\psi^2$ and $\psi_1$ ill defined in perturbation theory. In particular these will lead 
to additional logarithmic 
divergences in the loop amplitudes of the type \refb{eprop} with $L_0=0$ which
cannot be regarded as due to the collective modes and therefore cannot be removed by
the procedure described in \S\ref{stach}. This is already visible {\it e.g.} in the annulus amplitude
analyzed in \cite{1907.07688}.
To solve this problem we shall choose the
Lagrangian submanifold in this sector to be $\psi_1=0,\psi_2=0$, corresponding to the
original definition of fields and anti-fields. In this case the quadratic
term in the action, proportional to $\langle\Psi|Q_B|\Psi\rangle$, does depend on $\psi^1$ 
since $c_0|0\rangle$ is not BRST invariant, but does not depend on $\psi^2$ since $|0\rangle$
is BRST invariant. In fact once we integrate out the modes with $L_0>0$, for which we can 
use Siegel gauge condition without any problem, the only field in the expansion of 
$|\Psi_o\rangle$ multiplying ghost number 
$\ne 1$ state is $\psi^2$ and as a result the whole effective action becomes independent
of $\psi^2$ due to ghost number conservation of world-sheet correlators. 
Therefore the integration over $\psi^2$  factors out of the path integral, 
and its contribution can be absorbed into
the overall normalization factor ($\NN$ in \refb{er2}), 
leading to well defined perturbation theory. 

To understand this point better, it will be useful to recall the physical significance of
$\psi^2$ integration. Since $\psi^2$ is the coefficient of a ghost number 0 state 
$|0\rangle$ of the
open string, it represents a gauge transformation parameter, or equivalently the ghost
field corresponding to the gauge transformation parameter. BRST invariance of
$|0\rangle$ shows that the gauge transformation under consideration actually represents
a rigid gauge transformation. Indeed, this can be identified with the rigid $U(1)$ gauge 
transformation under which any open string stretching from 
the D-instanton to another
D-brane picks up a constant phase. 
Therefore integration over $\psi^2$ corresponds to division by
the volume of this $U(1)$ group. Since this is a constant factor, dropping this integral
just changes the overall normalization that can be absorbed in $\NN$.

However, as in the case of the collective coordinates discussed earlier, the open string 
gauge transformation parameter (equivalently ghost field) $\psi^2$ may be related
to the rigid $U(1)$ gauge transformation parameter $\theta$
by a complicated field dependent 
normalization. This can be detected by comparing the gauge transformation in open
string field theory generated by $\psi^2|0\rangle$ with the $U(1)$ gauge transformation
that gives a simple phase $e^{i\theta}$
for any open string stretching from the D-instanton to
another D-brane. If there is such a non-trivial field dependent normalization relating 
$\psi^2$ and $\theta$, we need
to change variable from $\psi^2$ to $\theta$, regarding both as grassmann odd
ghost fields, and then  drop the integration over $\theta$. The Jacobian associated 
with this change of variables will contribute to the
integration measure and therefore to the effective action as
in the case of integration over the open string zero mode $\phi$ discussed earlier.

To summarize, while in the $L_0\ne 0$ sector we continue to use the Siegel gauge condition
$b_0|\Psi_o\rangle=0$, in the $L_0=0$ sector we use the original definition of fields and 
anti-fields to define the Lagrangian submanifold, i.e.\ set the components of the open string field
with ghost number $\ge 2$ to zero. 
This removes the contribution due to the ghost zero modes from the propagator. 
Therefore,
in the perturbative amplitudes, we can remove the $q^{-1}$ terms in \refb{eprop} arising
due to ghost zero modes, just as we would remove the $q^{-1}$ terms arising from the
zero modes associated with the collective coordinate. However we now have to explicitly 
include the contribution from the $\psi^1$ propagator -- a contribution that is absent in the
usual world-sheet expression for the amplitude.

In order to evaluate the contribution to the amplitude due to the $\psi^1$ field, 
we shall need the form of the quadratic term in the action 
of the field $\psi^1$. For later use we shall compare this with the
quadratic term in the action for the tachyon field 
$\psi^0$ multiplying $c_1|0\rangle$. If we expand the open string field
$|\Psi_o\rangle$ of ghost number 1 as
\be
|\Psi_o\rangle = \psi^0 c_1|0\rangle + \psi^1 c_0|0\rangle +\cdots\, ,
\ee
then the quadratic term in the action is given by,
\be\label{esqu}
{1\over 2} \langle\Psi_o | Q_B |\Psi_o\rangle = {1\over 2} (\psi^0)^2 + 
(\psi^1)^2 +\cdots \, ,
\ee
where we have used $\{Q_B,c_0\}=2\, c_1c_{-1}$, the normalization convention
\be
\langle 0| c_1 c_0 c_{-1}|0\rangle =1\, ,
\ee
and the fact that the BPZ conjugation, that takes $|\Psi_o\rangle$ to
$\langle\Psi_o |$,  is generated 
by $z\to -1/z$. It follows from \refb{esqu} that the propagator of the 
tachyon $\psi^0$ is $-1$, which agrees with \refb{eprop}. \refb{esqu} also
shows that in the same normalization, the propagator for $\psi^1$ is $-1/2$.

\subsection{Summary of the algorithm} \label{ssum}

We can summarize the procedure for dealing with the divergences associated with open string degeneration 
as follows:
\begin{enumerate}
\item We compute amplitudes involving external on-shell closed string states and arbitrary number of insertions
of the on-shell open string zero 
modes $\phi$ associated with space-time
translation of the D-instanton. 
These amplitudes
can be expressed as integral over the moduli space
of punctured Riemann surfaces.
\item Near any degeneration where a pair of open string punctures are sewed together by a long strip, we
change variables so that the integral over the moduli space of Riemann surface is expressed as an
integral over
the parameters arising from string field theory. One
of them corresponds to the sewing parameter $q$ that comes
from Schwinger parameter representation of the propagator as given in \refb{eprop}. Others are integration
parameters that enter into the definition of the interaction vertex of string field theory.
In case of multiple degenerations where the Riemann
surface has several long strips, there are multiple sewing parameters $q_1,q_2,\cdots$, -- one for each open string propagator.
\item We expand the integrand as a power series in $q$. Using \refb{eprop},
an integral of the type $\int_0^1 dq q^{-1+h}$ is replaced
by $1/h$ both for $h>0$ and for $h<0$, as long as $h\ne 0$. For multiple degenerations, we do this for each
variable $q_i$.
\item A term in the integral of the form $\int_0^1 dq\, q^{-1}$ is set to zero. This corresponds to dropping
the path integral
over the Siegel gauge states with $L_0=0$. These include the zero modes $\phi$ associated with collective 
coordinates, as well as the zero modes  $\psi_1$, $\psi^2$ introduced in \refb{eexpand}. 
The justification
for dropping the path integral over $\psi_1$ and $\psi^2$ has been described in \S\ref{sghost}.
On the other hand, as discussed in \S\ref{stach},
the integration over the zero modes corresponding to collective coordinates is supposed to be carried out at
the end.
\item We need to compare the open string field theory gauge transformation generated
by $\psi^2|0\rangle$ with the simple $U(1)$ gauge transformation that gives a phase
$e^{i\theta}$ for any open string stretched from the D-instanton to another
D-brane. If $\psi^2$ and $\theta$ are related by field dependent normalization, we need 
to change variable from $\psi^2$ to $\theta$, regarding both as grassmann odd ghost
fields, and then
drop integration over $\theta$. The Jacobian associated with this change of variables
needs to be taken into account in all subsequent computations.
\item We now need to add the contribution from the intermediate $\psi^1$ state for each open string propagator. 
This
requires computing the relevant amplitude involving insertion of the states $c_0|0\rangle$ and 
multiplying it by
the $\psi^1$ propagator computed from \refb{esqu}. Since $c_0|0\rangle$ is not
a primary state, the result will depend on the choice of local coordinate system in which the corresponding vertex
operator is inserted. This information comes from string field theory.
\item The range $0\le q\le 1$ typically will span a subspace of the full moduli space
near a degeneration point. 
We can
carry out integration over the rest of the moduli space using the original variables, since there are no
divergences coming from this region. This corresponds to contribution from contact term vertices in string
field theory.
\item After computing the amplitudes by summing over all Feynman diagrams, we sum over all possible number
of insertions of
$\phi$ for a given set of external closed string states. This gives a function of the zero modes $\phi$.
We then integrate over $\phi$ to get the D-instanton contribution to the closed string amplitude. On general grounds we expect that there exists 
appropriate change of variables
from $\phi$ to the collective coordinates $\xi$ so that integration over $\phi$ 
reduces to a form proportional to $\int d\xi \, e^{ip.\xi}$, 
where $p$ is the total momentum carried by the 
external closed strings in an amplitude. This will recover the momentum conserving
delta function $\delta(p)$.
\end{enumerate}

\sectiono{Disk four point function} \label{swarm}

In this section we shall illustrate the breakdown of the Siegel gauge  
in the perturbative amplitudes.
We shall consider the $\phi$-$\phi$-$\phi$-$\phi$ four point function on the disk, where, for
definiteness, we shall choose $\phi$ to be the collective mode associated with 
the freedom of translating the D-instanton along the
(euclidean) time direction. 
Since this is a tree amplitude, and since $\psi_1$ is not a classical field, we shall not
see the need for dropping $\psi_1$ in the computation of this amplitude, but we shall
see the need for
including the contribution from the field $\psi^1$ separately. Furthermore, since the
$\phi$-$\phi$-$\phi$ three point coupling vanishes due to time reversal symmetry,
we shall not need to remove the contribution of the $\phi$ field in the internal 
propagator.

Amplitudes of this type have been analyzed previously
in \cite{0307019,1801.07607,1905.04958,1912.05463} for computing effective potential of massless
fields. However these computations used a particular form of string interaction vertex which has
an additional $Z_2$ symmetry known as twist symmetry, and
due to the use of twist symmetric three point vertex, they did not encounter
tree level breakdown of Siegel gauge for this amplitude. 
Nevertheless a field closely related to $\psi^1$
was discussed in \cite{1912.05463} (called $\vp_2$ there) in the context of heterotic string theory,
where it was observed that the coupling of this field to a pair
of massless fields vanishes due to a specific symmetry, 
and therefore this field does not appear as intermediate
state in the four point scattering amplitude. The role of twist symmetry in our analysis
will be discussed later in this section.

Even though our eventual interest is in computing amplitudes with one or more external
closed strings, a disk 4-point function with four $\phi$'s could arise as a disconnected
part of an amplitude with closed strings, {\it e.g.} the product of a disk one-point function
of a closed string and disk four point function of open strings. For this reason, it is
important to evaluate this amplitude. Our analysis will be independent of whether the other
coordinates are compact or non-compact, and they may even be replaced by a $c=25$
Liouville theory, representing two dimensional string theory.

Before we proceed with the actual computation, let us
discuss what result one should expect. The amplitude under consideration 
can be interpreted as the contribution to the $\phi^4$ term in the effective action
after integrating out all the open string modes other than those associated with
the collective coordinates. Since the effective action should be independent of the
collective coordinates, and since
the field $\phi$ is associated with the 
collective coordinate describing 
translation
of the D-instanton along the time direction,  the effective action should not depend 
on $\phi$. 
One might worry that $\phi$ may be related to the actual collective coordinate by
a field redefinition. However, since the effective action is altogether independent
of the collective coordinates, no field redefinition can produce a $\phi$ dependence of
the effective action.
Therefore we expect the four point amplitude to vanish.
This is what we shall now proceed to verify.

If we denote the world-sheet scalar field corresponding to the time
coordinate by $X$, then the unintegrated vertex operator for $\phi$ is $c\p X$ and the
integrated vertex operator is $\p X$. Then, up to a constant of proportionality, 
the amplitude is given by:
\be \label{eampli}
A=\int_0^1 dy \, \langle c\p X(0) \p X(y) c\p X(1) c\p X(\infty)\rangle
=\int_0^1 dy \left\{ {1\over y^2} +{1\over (1-y)^2} + 1
\right\}\, .
\ee
Note that we have included the contribution
from only the $0\le y\le 1$ region since the contribution from the other regions are related to
these by permutation of the external states accompanied by SL(2,R) transformations, 
and since all external states are identical, they
produce identical contributions.
The integral \refb{eampli} diverges from the $y=0$ and $y=1$ regions.  
In particular,
near $y=0$ and $y=1$ the integrand in \refb{eampli}
has double poles indicating tachyon propagation. 
This is expected, since 
the operator product of $\p X$ with itself generates an identity
operator. There is however no $\p X$ in the operator product of $\p X$ with itself,
therefore we do not need to subtract any collective coordinate 
contribution from the internal propagator. 
Our goal will be to
show
how to extract a finite result from \refb{eampli} 
following the procedure described in \S\ref{sdiv}.
As we shall see, we also
need to include the additional contribution due to the $\psi^1$ propagator that is not
present in the usual perturbative world-sheet amplitudes.

In order to proceed, we need to introduce the three point interaction
vertex of three open strings.
For external off-shell open string states $A_1$, $A_2$, $A_3$ the vertex takes the
form:
\be
\langle f_1\circ A_1(0)\, f_2\circ A_2(0)\, f_3\circ A_3(0)\rangle\, ,
\ee
where $f_1$, $f_2$ and $f_3$ are three conformal transformations, and $f\circ A$ is the
conformal transform of $A$ by $f$. We shall choose the functions $f_i$ such that $f_1(0)=0$,
$f_2(0)=1$, $f_3(0)=\infty$. We also take the vertex to be cyclically symmetric by requiring
that the transformation 
\be \label{ecyclic}
z\to {1\over 1-z}\, ,
\ee
cyclically permutes $f_1(w)$,
$f_2(w)$ and $f_3(w)$. This makes the vertex invariant under cyclic permutation of $A_1$,
$A_2$ and $A_3$.
In principle the vertex needs to be fully (anti-)symmetrized under the permutations of
$A_1$, $A_2$ and $A_3$. This can be done by averaging over the permutations of
the $A_i$'s, but since the vertices we shall use will always have two identical external
states, this will be automatic.

For simplicity we shall take the $f_i$'s to be SL(2,R) transformations. The
most general SL(2,R) transformations satisfying the desired properties are
labelled by a pair of parameters $\alpha$ and $\gamma$:
\be\label{ef1f2f3}
f_1(w_1) = {2 w_1 \over 2\alpha + w_1(1-\gamma)}, \quad
f_2(w_2) = {2\alpha + w_2(1-\gamma) \over 2\alpha - w_2(1+\gamma)}, \quad 
f_3(w_3) = -{2\alpha - w_3(1+\gamma) \over 2w_3}\, .
\ee
We shall take $\alpha$ to be a large number and ignore terms involving negative powers
of $\alpha$, although all final results are independent of
$\alpha$.
Denoting by $z$ the global coordinate on the upper half plane, we can identify $z$ with
$f_i(w_i)$ neat $w_i=0$. Inverting these relations we get:
\be \label{e3.2}
w_1 =\alpha \, {2\, z\over 2-z+\gamma\, z}, \quad w_2 
= 2\, \alpha\, {z-1\over z+1+\gamma(z-1)}, \quad 
w_3= 2\, \alpha\, {1\over 1+\gamma - 2\, z}\, .
\ee

We shall now consider the $s$, $t$ and $u$-channel diagrams obtained by gluing a pair 
of these 
vertices. Since the external states are all identical, it is sufficient to consider only
one of these diagrams  -- the others give identical contribution. We shall call this the contribution from the
amplitude with a propagator -- to be distinguished from the contribution from the four point
interaction vertex which does not have a propagator.
 For this we introduce two upper half planes labelled by $z,z'$ and local coordinates
$w_i$, $w_i'$ with $1\le i\le 3$ on each of these planes and make the identification:
\be 
w_2 w_2'=-q\, .
\ee
Using \refb{e3.2} we get:
\be
4\, \alpha^2 \, {z-1\over z+1+\gamma(z-1)}\, {z'-1\over z'+1+\gamma(z'-1)} = -q\, .
\ee
The four external punctures of the four point function are located at $z=0,\infty$ and
$z'=0,\infty$. In the $z$ plane 
they are located at
\be \label{e3.8}
z=\infty, \quad z=0,
\quad z'=\infty \ \Rightarrow \ z= {4\alpha^2 + (\gamma^2-1)q\over 4\alpha^2 + (1+\gamma)^2 q}, \quad z'=0 \ \Rightarrow \ z= {4\, \alpha^2+ (1-\gamma)^2\, q\over
4\alpha^2 +(\gamma^2-1)q}\, .
\ee

We shall now make an SL(2,R) transformation to bring three of the punctures at 0, 1 and
$\infty$, keeping the fourth puncture between 0 and 1.
Under SL(2,R) transformation
\be 
\hat z = z \, {4\alpha^2 +(\gamma^2-1)q\over 4\, \alpha^2+ (1-\gamma)^2\, q}\, ,
\ee
the punctures are located at:
\be \label{e3.7}
\hat z=\infty, \quad \hat z=0, \quad \hat z = 
{4\alpha^2 +(\gamma^2-1)q\over 4\, \alpha^2+ (1-\gamma)^2\, q}\,
{4\alpha^2 + (\gamma^2-1)q\over 4\alpha^2 + (1+\gamma)^2 q}
= 1 - {q\over \alpha^2}+{(1+\gamma^2) q^2 \over 2\alpha^4}+\OO({q^3\over
\alpha^6}), \quad \hat z=1 \, .
\ee
On the other hand under SL(2,R) transformation 
\be
\tilde z = 1 - {1\over z} \, {4\alpha^2 +(\gamma^2-1)q\over 4\, \alpha^2+ (1+\gamma)^2\, q}\, ,
\ee
the punctures are located at
\be \label{e3.9}
\tilde z =1, \quad \tilde z = \infty, \quad \tilde z = 0, \quad \tilde z = 1 - 
{4\alpha^2 +(\gamma^2-1)q\over 4\, \alpha^2+ (1+\gamma)^2\, q}
\, {4\alpha^2 +(\gamma^2-1)q\over 4\, \alpha^2+ (1-\gamma)^2\, q}
= {q\over \alpha^2}-{(1+\gamma^2) q^2 \over 2\alpha^4}+\OO({q^3\over \alpha^6})\, .
\ee
Eqs.\refb{e3.8}, \refb{e3.7} and \refb{e3.9} give equivalent representations of the puncture
locations for the Feynman diagrams with a propagator.

Let us now turn to the amplitude \refb{eampli}.
For analyzing the singular contribution to \refb{eampli} from near $y=1$,
we denote
the location of the third puncture in  \refb{e3.7} by $y$. This gives:
\be 
1-y = {q\over \alpha^2} - {(1+\gamma^2) q^2 \over 2\alpha^4}+\OO(q^3/\alpha^6)\, .
\ee
The range $0\le q\le 1$ corresponds to:
\be \label{erange1}
1 - {1\over \alpha^2} + {(1+\gamma^2)  \over 2\alpha^4}+\OO(\alpha^{-6})\le y\le 1\, .
\ee
We also have
\be
(1-y)^{-2} dy = -\alpha^2 \, q^{-2} \, dq + \OO(\alpha^{-2})\, .
\ee
Our strategy will be to change variable from $y$ to $q$ in the range \refb{erange1} and
interpret the contribution from this region as coming from Feynman diagram with a
propagator, with the divergence in the integrand as due to tachyon propagating along the
internal propagator.\footnote{Physically, 
the contribution from the region \refb{erange1} may be regarded as coming from
the $s$-channel diagram,  the contribution from the 
region \refb{erange2} may be regarded as coming 
from the $t$-channel diagram, and the contribution from
the rest of the region of $y$-integration
may be interpreted as coming from the four point contact interaction.
For the particular cyclic ordering we have chosen, there is no $u$-channel 
diagram.}
Therefore we write
\be \label{e3.14}
\int_0^1 dy\, (1-y)^{-2} = \int_0^{1 - {1\over \alpha^2} + {(1+\gamma^2)  
\over 2\alpha^4}+ \OO(\alpha^{-6})} \, dy\, (1-y)^{-2} 
+ \alpha^{2} \int_0^1 {dq\over q^2} + \OO(\alpha^{-2})\, .
\ee
Using \refb{eprop}, we get the replacement rule:
\be 
\int_0^1 dq \, q^{-2} \quad \Rightarrow \quad -1\, .
\ee
Substituting this into \refb{e3.14} we get:
\be \label{efirint}
\int_0^1 dy\, (1-y)^{-2} = -1 + \left\{  {1\over \alpha^2} - 
{(1+\gamma^2)  \over 2\alpha^4}\right\}^{-1} -\alpha^2 + \OO(\alpha^{-2})
= {\gamma^2-1\over 2} + \OO(\alpha^{-2})\, .
\ee
Such change of variable should also be done for the $y^{-1}$ and 
$1$ terms, but since they are not singular at $y=1$, the change of variable will have no effect on
the value of the integral.

Similarly for evaluating the integral $\int_0^1 dy\, y^{-2}$, which is singular near
$y=0$, we denote the last puncture in \refb{e3.9} by $y$. This gives
\be
y = 1 - 
{4\alpha^2 +(\gamma^2-1)q\over 4\, \alpha^2+ (1+\gamma)^2\, q}
\, {4\alpha^2 +(\gamma^2-1)q\over 4\, \alpha^2+ (1-\gamma)^2\, q}
= {q\over \alpha^2}-{(1+\gamma^2) q^2 \over 2\alpha^4}+\OO({q^3\over \alpha^6})\, .
\ee
In this case the range $0\le q\le 1$ corresponds to
\be \label{erange2}
0\le y\le  {1\over \alpha^2}-{(1+\gamma^2)  \over 2\alpha^4}+\OO(\alpha^{-6})\, .
\ee
Also we have
\be
dy \, y^{-2} = \alpha^2 \, q^{-2} \, dq+ \OO(\alpha^{-2})\, .
\ee
Following the same strategy as before, we write
\be\label{e3.19}
\int_0^1 dy\, y^{-2} =\alpha^2 \int_0^1 dq\, q^{-2} + 
\int_{ {1\over \alpha^2}-{(1+\gamma^2)  \over 2\alpha^4}+\OO(\alpha^{-6})}^1 dy\, y^{-2}
+\OO(\alpha^{-2})\, .
\ee
After  using the replacement
rule \refb{eprop} for the first term, we get,
\be \label{esecint}
\int_0^1 dy\, y^{-2} = -\alpha^{2} - 1 
+ \left\{{1\over \alpha^2}-{(1+\gamma^2)  \over 2\alpha^4}\right\}^{-1} +\OO(\alpha^{-2}) 
= {\gamma^2-1\over 2}  +\OO(\alpha^{-2})\, .
\ee

Finally we also have the non-singular integral
\be \label{ethiint}
\int_0^1 dy = 1\, .
\ee
Adding \refb{efirint}, \refb{esecint} and \refb{ethiint}, taking
$\alpha\to\infty$ limit, and using \refb{eampli}, we get
\be \label{eanaive}
A = \gamma^2\, .
\ee

This however is not the full story. As argued in \S\ref{sghost}, we also need to 
add to this the contribution due to the $\psi^1$ exchange.
From
\refb{esqu} we see that this contribution is similar to the tachyon exchange contribution
$A_{\psi^0}$,
except for two differences. First, due to the absence of the 1/2 factor multiplying the
$(\psi^1)^2$ term in \refb{esqu}, the $\psi^1$ propagator is 1/2 of the tachyon propagator and
we shall have a factor of 1/2. Second, while $A_{\psi^0}$
will be proportional to the square of the $\phi$-$\phi$-$\psi^0$ three point coupling
$C_{\phi\phi\psi^0}$, the
$\psi^1$ exchange contribution $A_{\psi^1}$
will be proportional to the square of the 
$\phi$-$\phi$-$\psi^1$ three point coupling $C_{\phi\phi\psi^1}$. 
Therefore we have:
\be\label{eapsi1}
A_{\psi^1}= {1\over 2} \, A_{\psi^0} \, (C_{\phi\phi\psi^1}/C_{\phi\phi\psi^0})^2 \, .
\ee
The total tachyon exchange contribution to the amplitude $A_{\psi^0}$ is given by the terms in 
\refb{e3.14} and \refb{e3.19} from the $\int_0^1 dq\, q^{-2}$ part of the integral. 
Using the
replacement rule \refb{eprop}, we get:
\be \label{eat}
A_{\psi^0}= -2\, \alpha^2 \, .
\ee
The three point coupling with a pair of on-shell fields $\phi$, with vertex
operators $c\,\p X$ inserted at 0 and $\infty$, and an off-shell field with vertex
operator $V$ inserted at 1, is given by 
\be
\langle c\p X(0) \, f_2\circ V(0)\, c\p X(\infty)\rangle\, ,
\ee
where $f_2$ has been given in \refb{ef1f2f3}.
For the tachyon $V=c$ and we have
\be \label{e428}
f_2\circ c(0) = f_2'(0)^{-1} \, c(f_2(0)) = \alpha\, c(1)\, .
\ee
Furthermore, SL(2,R) invariance gives,
\be \label{eccc}
\langle c\p X(0) \, c(z) \, c\p X(\infty)\rangle = C\, z\, ,
\ee
where the normalization constant $C$ depends on the normalization and signature
of $X$. Using \refb{e428} and \refb{eccc} we get,
\be \label{eppp0}
C_{\phi\phi\psi^0} = \langle c\p X(0) \, \alpha c(1) \, c\p X(\infty)\rangle = C\, \alpha\, .
\ee
On the other hand, for $\psi^1$, $V=\p c$ and we have
\be
f_2\circ \p c(0) = \p c(f_2(0)) - {f_2''(0)\over f_2'(0)^2} \, c(f_2(0)) = \p c(1) - (1+\gamma) c(1)\, .
\ee
Using \refb{eccc} we get
\be \label{eppp1}
C_{\phi\phi\psi^1} = \langle c\p X(0) \, \{\p c(1) - (1+\gamma) c(1)\} \, c\p X(\infty)\rangle
= -C\, \gamma\, .
\ee
Substituting \refb{eat}, \refb{eppp0} and \refb{eppp1} into \refb{eapsi1}, we get
\be \label{epsifin}
A_{\psi^1} ={1\over 2} (-2\, \alpha^2)
\, {\gamma^2\over \alpha^2} = -\gamma^2\, .
\ee

Adding \refb{epsifin} 
to \refb{eanaive} we get the net contribution to the $\phi$-$\phi$-$\phi$-$\phi$ four
point function:
\be
A_{\phi\phi\phi\phi} = A + A_{\psi^1} = \gamma^2 -\gamma^2=0\, .
\ee
This is consistent with the 
identification of $\phi$ with the collective coordinate up to field redefinition.

Note that if we had set $\gamma=0$, then $C_{\phi\phi\psi^1}$ would have vanished, and as
a result there would be no $\psi^1$ exchange contribution. This is related to the fact that
for $\gamma=0$ the functions $f_1$, $f_2$ and $f_3$ defined in \refb{ef1f2f3}
are not only
cyclically permuted under the SL(2,R) transformation \refb{ecyclic}, but also has 
full permutation
symmetry, up to a change in the sign of the arguments $w_i$. 
For example the $z\to 1-z$ transformation exchanges
$w_1\leftrightarrow -w_2$ and sends $w_3$ to $-w_3$.
This leads to a `twist symmetric'
three point vertex where the twist symmetry is 
a $Z_2$ symmetry that assigns quantum number 
$(-1)^{h+1}$ to a component field that multiplies a world-sheet state of $L_0$ eigenvalue 
$h$\cite{kost,9705038}. 
Under this symmetry transformation the tachyon $\psi^0$ is even since it multiplies the state
$c_1|0\rangle$ of $L_0$ eigenvalue $-1$, 
the zero mode field $\phi$ is odd and the field $\psi^1$ is odd.  Therefore 
$C_{\phi\phi\psi^1}$ vanishes but $C_{\phi\phi\psi^0}$ does not vanish. For this reason,
it is convenient to use a twist symmetric vertex by setting 
$\gamma=0$, since this will avoid propagating $\psi^1$ in tree amplitudes.
However $\psi^1$ will still propagate in the loop and its contribution need to
be included separately.

\sectiono{Superstrings} \label{ssuper}

The problem with zero modes of D-instantons associated with world-sheet ghosts in not
unique to bosonic string theory. If we denote by $|-1\rangle$ the NS sector vacuum with
picture number $-1$, then the analog of the expansion \refb{eexpand}
for the NS sector string field can be
written as\footnote{In the Ramond sector, the full BV formalism requires a doubling of the
string fields\cite{1508.05387,1602.02582,1602.02583,1907.10632}, 
but this can be avoided in the NS sector by identifying the two sets of
string fields. For simplicity of notation, this is the approach we are adopting here.}
\be\label{eexpandsusy}
\psi^1 \, \beta_{-1/2} c_0 c_1  |-1\rangle +\psi^2\, \beta_{-1/2} c_1 |-1\rangle
+ \psi_1 \, \gamma_{-1/2} c_1 |-1\rangle + \psi_2 \, \gamma_{-1/2} c_0 c_1 |-1\rangle
\, ,
\ee
where $\beta_n$ and $\gamma_n$ are the modes of the superghost fields $\beta,\gamma$.
Based on ghost number counting of states, we shall regard $\psi^1$ and $\psi^2$ as fields
and $\psi_1$ and $\psi_2$ as their corresponding anti-fields. Siegel gauge fixing corresponds
to choosing a Lagrangian submanifold that sets $\psi^1$ and $\psi_2$ to zero. In the resulting
gauge fixed action, $\psi_1$ and $\psi^2$ appear as zero modes, causing perturbation theory
to diverge. As in the case of bosonic string theory, the remedy is to choose a different
Lagrangian submanifold by setting $\psi_1$ and $\psi_2$ to zero. In this case the mode
$\psi^2$ decouples from the action by ghost number conservation. On the other hand 
the mode $\psi^1$
has a non-zero kinetic term, and its contribution must be included separately in the
perturbation theory.

A similar analysis can be carried out in the Ramond sector using the identification of
fields and anti-fields described in \cite{1907.10632}.

\bigskip

\noindent {\bf Acknowledgement:} I wish to thank Carlo Maccaferri, Xi Yin and
Barton Zwiebach for useful discussions.
This work was
supported in part by the 
J. C. Bose fellowship of 
the Department of Science and Technology, India and the Infosys chair professorship.

\end{document}